# Physical origin of the mechanochemical coupling at interfaces

Zhuohan Li, Izabela Szlufarska

Department of Materials Science & Engineering
University of Wisconsin-Madison

**Abstract:**
We used density functional theory (DFT) calculations to investigate the physical origin of the mechano-chemical response of materials interfaces. Our results show that the mechano-chemical response can be decomposed into the contribution from the interface itself (deformation of interfacial bonds) and a contribution from the underlying solid. The relative contributions depend on the stiffness of these regions and the contact geometry, which affects the stress distribution within the bulk region. We demonstrate that, contrary to what is commonly assumed, the contribution to the activation volume from the elastic deformation of the surrounding bulk is significant and, in some case, may be dominant. We also show that the activation volume and the mechanochemical response of interfaces should be finite due to the effects on the stiffness and stress distribution within the near-surface bulk region. Our results indicate that the large range of activation volumes measured in the previous experiments even for the same material system might originate from the different degrees of contributions probed from the bulk vs. interface.

**Main Text:**

The term mechanochemistry refers to the coupling between chemical reactions and the mechanical strain in the system [1–3]. The mechanochemical coupling has been generally described by phenomenological theories, such as the Eyring [4], Bell [5], and Zhurkov [6] models. These theories were developed from different scientific perspectives, however, many of them share a similar physical foundation and are described by a similar mathematical expression $E_a = E_{a,0} - \Delta V P$. Here, $P$ is the stress acting on the system, $E_{a,0}$ is the stress-free energy barrier, $E_a$ is the effective energy barrier under the mechanical force, and $\Delta V$ is the proportionality constant. This equation shows that the free energy difference that needs to be overcome during a chemical reaction can be affected by the mechanical work done on the system, and the magnitude of the response to the applied stress depends on the constant $\Delta V$. $\Delta V$ has the units of volume, and thus it is referred to as the activation volume. However, the physical meaning of $\Delta V$ has been elusive [3,7,8]. The activation volume has been often considered to be related to the local deformation of the chemical bonds during the reaction [9,10]. However, the range of experimentally measured activation volumes can be quite large even for the same material system. For example, for tribochemical wear of silicon tips, the range of activation volumes can be as large as ~6.7-115 Å$^3$ [9,11–13]. With such a large range, the meaning of "local deformation" becomes obscure, and the simple physical picture of $\Delta V$ corresponding approximately to the size of the space occupied by molecular groups involved in the chemical reaction might not be sufficient.

Mechanochemical reactions are particularly important for the chemically active interface that is subject to compression and/or shear [9–23]. Understanding the nature of the activation volume and of the factors that control it is critical for interpretation of chemical reactions at interfaces, with numerous applications such as rock friction [24], micro/nano devices [25], wafer bonding [26], etc. Here, we demonstrate that for interfacial chemical reactions, the mechanochemical coupling is not solely due to the local deformation of chemical bonds right at the reaction site, as has been often assumed. Instead, we find that there is a significant contribution to this coupling from the deformation of the surrounding bulk material. Our conclusions are based on a series of density functional theory (DFT) calculations of interfacial chemical bonding reactions.

Here, we chose silica as a model material, as its surface properties are relevant to a number of fields, such as electronics and geophysics, as mentioned above [12,13,16–19,24–27]. We selected some of the lowest-energy surfaces of silica polymorphs, i.e., (001) and (111) surfaces of β-cristobalite, and (0001) surface of α-quartz. The three types of surfaces will be henceforth referred to C(001), C(111), and Q(0001), respectively. All surfaces were fully terminated with hydroxyl groups (-OH), forming surface silanols Si-OH. Both, C(001) and Q(0001) surfaces have geminals on the surfaces, where two -OH groups are attached to one surface Si atom. In contrast, the surface Si atoms on the C(111) are terminated with isolated single silanol groups. The different combinations of bulk structures and surface chemistries are chosen in this study to probe the generality of our conclusions. DFT calculations of silica surfaces were performed using the Vienna Ab initio Simulation Package (VASP) code [28]. Simulation details are provided in Supplemental Material [39].

Mechanochemical behavior of the interfacial reactions was investigated by calculating the reaction energy $\Delta E$ of siloxane bond formation (Si-OH + Si-OH = Si-O-Si + $H_2O$) [29] between two opposing slabs at different indentation depths. The relaxed configurations before and after the bond formation are referred to as initial and final structures, respectively. An example of C(001) is shown in Figs. 1 (a) and 1 (b). Traditionally, the activation volume has been introduced in the expression for the pressure-dependent energy barrier. However, the change in the reaction energy is often found to be linear with the change in the energy barrier, as described by the Brønsted-Evans-Polanyi (BEP) relation [30], which was also verified by our DFT calculations. Since DFT calculations of the reaction energies are computationally less expensive, we calculate reaction energies for all the interfaces considered in our study. In Supplemental Material [39], we show additional results for pressure-dependence of the energy barrier on one of the interfaces, i.e., C(001), where a similar trend to that of the reaction energy have been found. In the following, we will use the term "activation volume" to refer to the pressure dependence of both, the reaction energy and the energy barrier.

When a bond forms, atomic configuration changes not only right at the reaction site, but it also leads to deformation of silica tetrahedra in the surrounding bulk [18] and an associated change in the elastic energy of the bulk. To calculate the bulk contribution $\Delta E_{bulk}$, we first relax the entire system, where we can obtain the total reaction energy $\Delta E_{total}$, then we remove the interfacial atoms (interfacial -OH groups) as shown in Fig. 1 (c), and finally we perform a single-point calculation without further structural relaxation. The bulk contributions from the upper and the lower slabs are calculated separately (by placing each slab in a vacuum) with a dipole correction applied along the z-direction. Interfacial contribution $\Delta E_{interface}$ is then obtained by subtracting the bulk contribution from $\Delta E_{total}$.

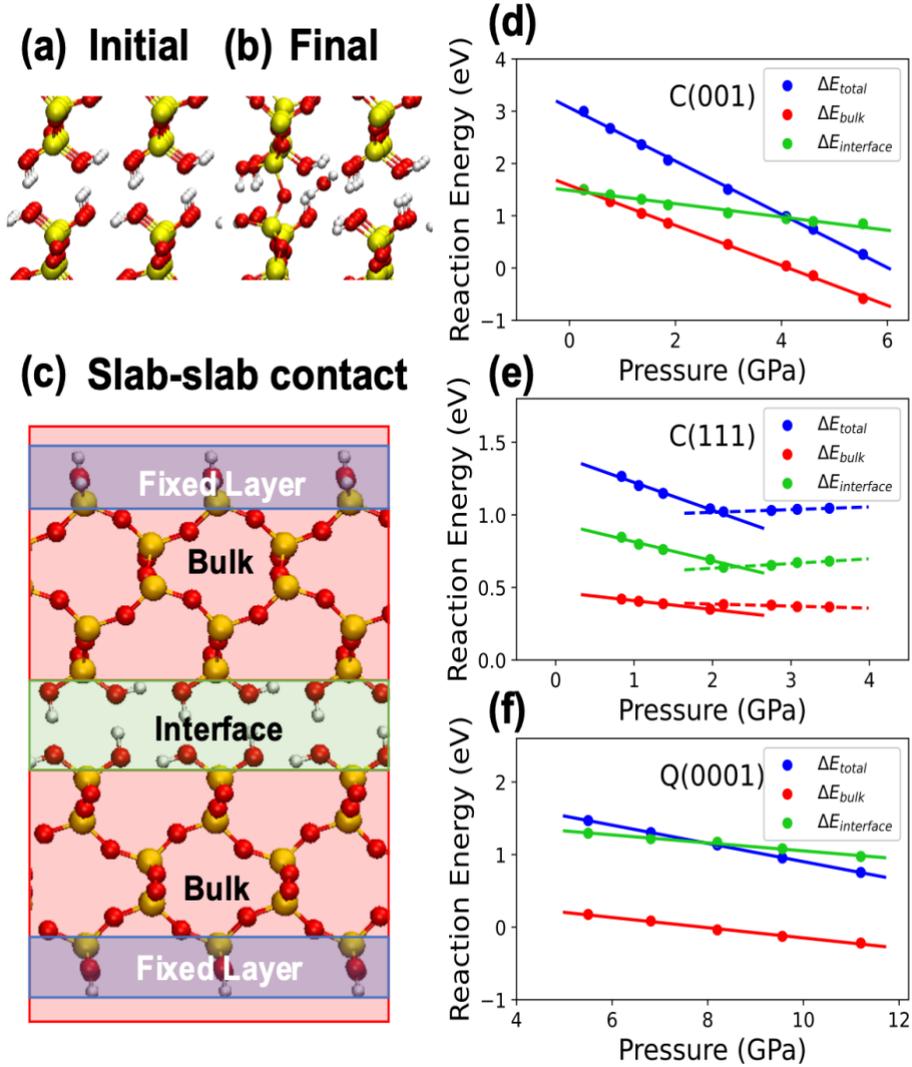

FIG. 1. Atoms in the initial (a) and the final (b) structure for the C(001) interface. (c) A supercell containing the C(001) interface before reaction. The outermost Si, O, and H atoms of silanol groups Si-OH are kept fixed during the geometric relaxation. The bulk part of the system is defined as all the atoms, including the atoms in the fixed layers, without the hydroxyl groups (O and H atoms) that are within the contact interface. The bulk, interface, and fixed layers in the final structure are defined in the same way. Yellow: Si, Red: O, White: H. [(d)-(f)] Reaction energy vs. normal contact pressure for C(001), C(111), and Q(0001), respectively. Blue: total reaction energy $\Delta E_{total}$, red: bulk contribution $\Delta E_{bulk}$, green: interfacial contribution $\Delta E_{interface}$. For all figures, solid and dashed lines are obtained by a linear regression to the data points marked as circles.

As shown in Figs. 1 (d)-(f), the total reaction energy $\Delta E_{total}$ (blue) shows a linear dependence on the contact pressure, at least within a certain range of pressures, which agrees with the Eyring model. In the case of the C(001) and Q(0001) interface, $\Delta E_{total}$ decreases with an increasing contact pressure in the entire pressure regime, as shown in Fig. 1 (d) and (f). The absolute value of the slopes corresponds to the activation volumes $\Delta V_{total}$ of the reaction energy on C(001) and

Q(0001), respectively. Fig. 1 (e) shows that the C(111) interface has two different activation volumes depending on the range of contact pressures considered. In the low-pressure region (< 2.0 GPa, denoted as C(111)$_{low}$), the reaction energy decreases with pressure, which is the same as for the other two interfaces. In the high-pressure regime (> 2.0 GPa, denoted as C(111)$_{high}$), there is still a linear relationship between the reaction energy and the contact pressure, but the slope of the line becomes much smaller and $\Delta E_{total}$ actually slightly increases with the contact pressure. One possible reason for the existence of the two different activation volumes for C(111) is the reconstruction of the interfacial Hydrogen bond (H-bond) network when the contact pressure reaches a transition pressure. As shown in Figs. 2 (a) and 2 (b), there is a clear change in the interfacial H-bond network structure below and above the pressure of ~2 GPa. This kind of interfacial structural change is absent in the case of C(001) and Q(0001) in our calculations.

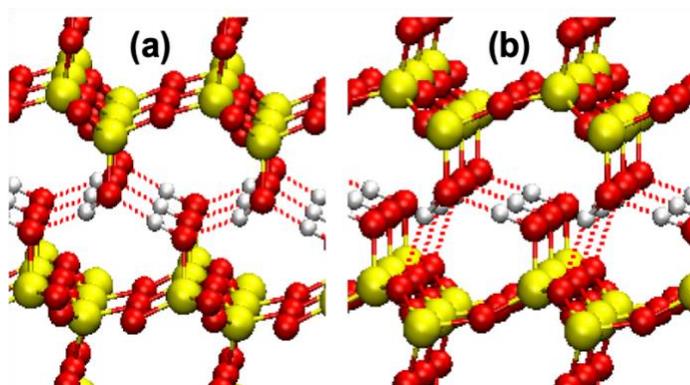

*FIG 2. H-bond network at the C(111) interface at (a) low contact pressures (< 2.0 GPa) and (b) high contact pressures (> 2.0 GPa). ) Yellow: Si, Red: O, White: H. Hydrogen bond is plotted as dashed red lines. To define H bonds, we used the cutoff of 3.3 Å for O-O between adjacent silanols [31], and the cutoff of 145º for O-H-O angles [32].*

Another important observation from our studies is that there is a non-negligible (and in some cases dominant) contribution from the bulk region to the mechanochemical response. As shown in Fig. 1 (d) for C(001), the major contribution to the change of the total reaction energy $\Delta E_{total}$ with pressure comes from the change in the bulk deformation energy ($\Delta E_{bulk}$), whereas the change in the interfacial contributions ($\Delta E_{interface}$) is relatively insignificant (i.e., $|\Delta V_{bulk}| > |\Delta V_{interface}|$). On the other hand, for C(111) and Q(0001) (see Figs. 1 (e) and 1 (f), respectively), the contributions from the bulk and from the interface are more comparable (i.e., $|\Delta V_{bulk}| \approx |\Delta V_{interface}|$). In either case, it is surprising that the energy associated with a deformation of the molecules in the interfacial region (where a chemical reactions actually occur), is not necessarily the dominant factor in how the energy of that chemical reaction depends on pressure, i.e., in the mechanochemical response of the system.

The relative contributions from the bulk and the interface to the mechanochemical response of an interfacial chemical bonding reaction can be described by a first-order perturbation theory. This theory was originally developed elsewhere [33], but here it is adapted to describe the interfacial chemical bonding reaction and mathematically formulate the contributions from the bulk and the interface. A detailed derivation is provided in Supplemental Material [39]. Briefly, the activation volume for the interfacial chemical bonding reaction can be expressed as $\Delta V = -\Delta F(s)/k^*(0)$, where $s$ represents a state along the reaction path. The variable $s$ is equal to 0 in the initial state, 1 in the final state, and $0 < s < 1$ in the transition state. $\Delta F(s)$ is the change in the contact force along z-direction. This change is between states 0 and $s$ and can be written as $\Delta F(s) = F(s) - F(0)$. $k^*(0)$ is the contact stiffness per unit contact area at the initial state, i.e., $k^*(0)A = k(0)$, where $A$ is the nominal contact area of the interface.

The above expression for the activation volume represents the total activation volume $\Delta V_{total}$, and it can be further decomposed into the contributions from bulk, $\Delta V_{bulk}$, and the interface, $\Delta V_{interface}$, i.e., $\Delta V_{total} = \Delta V_{bulk} + \Delta V_{interface} = -\Delta F(s)/k_b^*(0) - \Delta F(s)/k_i^*(0)$, where $k_b^*$ and $k_i^*$ are the stiffness of the bulk and the interface per unit contact area, respectively. This decomposition is possible under the simplifying assumptions that (1) the contacting materials can be treated as multiple springs in series along the z-direction, and (2) the stress field is uniform across the slab and consequently, $\Delta F(s)$ is assumed to be the same for the bulk and the interface for a given interface. Both $\Delta F(s)$ and $k(0)$ can be obtained from DFT calculations (see Supplemental Material [39]).

In Fig. 3, we plot $\Delta V$ estimated by the above first-order perturbation analysis against the values of $\Delta V$ determined as the slopes of the linear fits to DFT data shown in Fig. 1 [(d)-(f)]. The linear fit is quite good with $R^2 = 0.972$ and the mean absolute error (MAE) = 3.29 Å$^3$. The error could possibly arise from ignoring the shear components of the stress tensor and assuming the stress field to be uniform over the near-interface region. These approximations are expected to affect the accuracy of the prediction of the activation volume. For example, the local stress can be different in the bulk region and at the interface, which is not considered here. Nevertheless, the accuracy of the predictions for the total activation volume is quite high, where almost perfect correlation is obtained between the DFT calculations and the theory (see blue markers in Fig. 3).

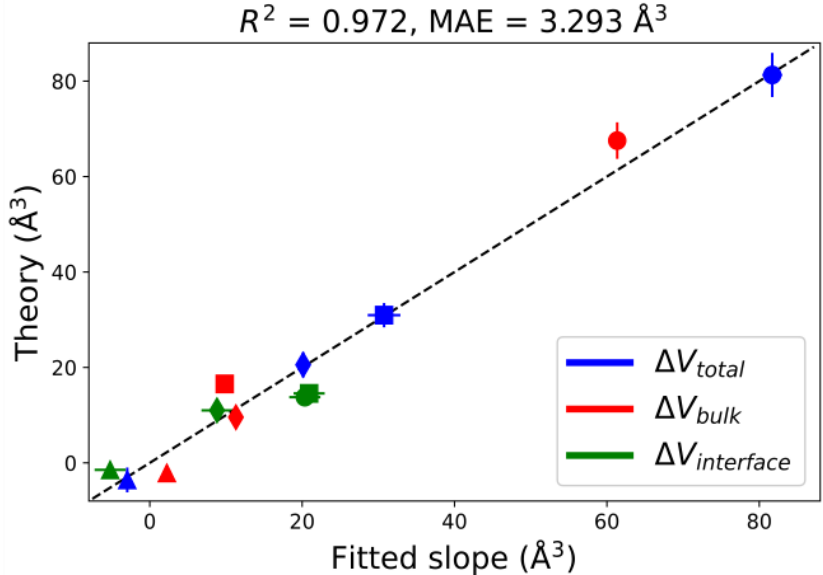

*Fig. 3 Comparison of the activation volumes obtained by first-order perturbation theory with the slopes of linear fits to DFT data shown in Fig. 1. Circles: C(001), squares: C(111)$_{low}$, triangles: C(111)$_{high}$, and diamonds: Q(0001). Blue, red, and green colors correspond to the total $\Delta V_{total}$, bulk $\Delta V_{bulk}$, and interfacial $\Delta V_{interface}$ contributions, respectively. The error bars for fitted slopes are the standard error of the slopes, and the error bars in the theory are obtained from propagation of standard errors. The data points should lie on the dashed line in the case of a perfect correlation.*

According to the above theory, the activation volume will be infinite if we assume 1) a constant elastic modulus in the bulk region, and 2) a uniform stress field inside the bulk (since the activation volume is an extensive variable). The second assumption was used in our first-order perturbation theory. However, the true activation volume in real material contacts should still be finite and we have proposed two near-surface phenomena that are responsible for that.

The first phenomenon is the surface effect on the elastic modulus — we found that the stiffness of our silica slab is much lower than what was reported for bulk silica. Our DFT calculations show that the $C_{33}$ component of the stiffness tensor of the bulk part of our C(001) slab is only 45.8 GPa, whereas it is 186 GPa for the bulk unit-cell with periodic boundary conditions in all three directions. The latter is close to the previously reported value for β-cristobalite, i.e., ~194-196 GPa [34]. On the other hand, it was also reported that 2D silica (a chemically stable bilayer of $SiO_2$) has Young's modulus of 43 GPa, which is much smaller compared to 72 GPa of silica glass (fused silica) [35].

We hypothesize that the difference in the stiffness is due to the surface effect, i.e., the structural relaxation of tetrahedra in the near surface region, which leads to different bond length and bond angle distributions from those inside the bulk. In order to confirm whether the soft slab is an artifact that arises from having too few atomic layers in our simulations, we relaxed a thicker C(001) supercell that had a thickness of 20 unit cells (layers). The bottom 2 layers (the 19$^{th}$ and 20$^{th}$ layer) are fixed in the perfect crystal structure configuration, which is obtained independently by optimizing a bulk unit cell with periodic boundary conditions in all spatial directions. In Fig. 4 we plot the Si-O bond lengths and O-Si-O angle distributions in different layers in the thick 20-layer slab, as well as the corresponding distributions obtained for the thin slab we used for other calculations. The bond length and bond angle distributions of the thin slab are found to be quite close to the distributions of the 1$^{st}$ unit cell layer of the large supercell. Further away from the surface, the bond length and bond angle distributions gradually become more similar to the distributions in the bottom fixed layer. Since our DFT calculations show that the thin silica slab has a lower stiffness than the bulk silica, the gradual change in the crystal structures implies a gradual increase in the stiffness along the normal direction from the surface into the bulk. Based on the first-order perturbation theory, this means that a softer near-surface region should have a much larger contribution to the overall activation volume than the region deep inside the bulk. As shown in Fig. 4, the bond length and bond angle distributions do not truly converge to the bulk values within 20 layers. This trend means that the structural change, and therefore the stiffness change, in the near-surface region should extend even further than 20 unit-cell layer thickness for C(001).

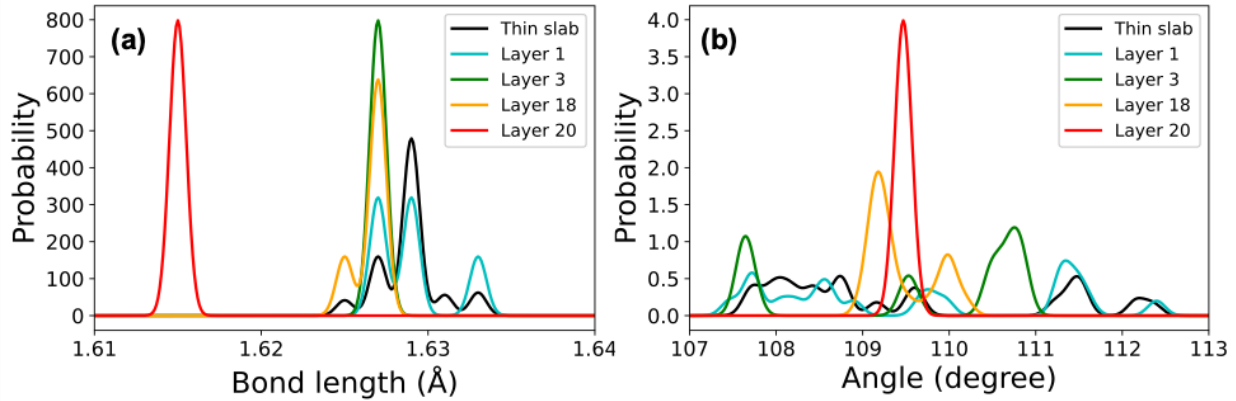

*Fig. 4 Si-O bond length and O-Si-O bond angle distributions of the thin (1 unit cell) slab and the thick (20 unit cells) slab of C(001). Si-O bond lengths and O-Si-O bond angles that involve surface Si-OH groups are ignored. Configuration of layer 20 in the thick slab was set to the configuration obtained from bulk unit cell optimization, where Si-O bond length = 1.614 Å, and O-Si-O angle = 109.47°. Here, the distributions correspond to the kernel density estimate (KDE) of the histograms with a bin width of 0.002 Å and 0.002° for bond length and bond angle, respectively. The band widths used for KDE are 0.0005 Å in (a) and 0.1° in (b),*

One other factor that affects the activation volume besides the stiffness is the stress field within the bulk, which appears in the numerator of the expression for the activation volume obtained from first-order perturbation theory. So far, we have assumed a uniform stress field in the bulk, but it might not be the case for a real rough contact interface, where the elastic energy is mainly stored in the near-surface regions and the highest stress will be in the vicinity of the contacting asperities [36]. This physical situation will result in the decay of $\Delta F(s)$ from the surface into the bulk. As a result, the local activation volume will also decay as a function of distance from the surface. To estimate the order of magnitude of this effect, we conducted simplified calculations based on the Hertzian contacts [37], where we considered cases of materials with fixed Young's modulus between 50 GPa (soft) and 200 GPa (stiff) and single asperity (or tip) radii from 20 nm (sharp) to 2.5 µm (blunt). Details of calculations are provided in Supplemental Material [39]. We found that for these conditions, the local mechanochemical response decays to a negligible value within ~5-200 nm from the surface. That means that the mechanochemical response due to the elastic deformation should still be dominated by the near-surface region and therefore it will be finite. We also roughly estimated the ratio of the mechanochemical response $\Delta VP$ from the bulk (near-surface region) and the interface region. We found this ratio to be ~1.5 and ~ 203.6 for a stiff sharp tip and a soft blunt tip, respectively. In this estimate we assumed the interfacial stiffness to be the average value of the different silica-silica interfaces considered earlier in our study. This result implies that the contribution from the bulk region to $\Delta VP$ cannot be ignored and that it depends on the contact stiffness and geometry.

That said, in some cases the interface contribution to stiffness can still be dominant. For example, for tribochemical polymerization [4,22], the activation volume (~10 Å³) was considered to be related to the shear-induced deformation of molecules adsorbed at the interface. This conclusion is in fact consistent with our theory. For the interfaces in Refs. [4,22], the adsorbed molecules have a large number of degrees of freedom for deformation as compared to the atoms in the solid substrate. Therefore, the effective stiffness of those molecules is likely much lower than the stiffness of the substrate, which would result in a larger contribution from the interface to

Δ*V*. Similar observations were also obtained in hydrostatic-pressure-driven redox reactions in metal–organic chalcogenides [38]. The authors showed that for copper(I) m-carborane-9-thiolate (Cu-S-M9) crystals, the charge reduction from Cu(I) to Cu(0) is induced by the anisotropic deformation of the "soft" $Cu_4S_4$ mechanophore and relative motions of the rigid M9 ligands that surround the $Cu_4S_4$ core.

In conclusion, we have demonstrated that a significant contribution to the mechanochemical coupling at the interface can arise from deformation of the bulk and this contribution will depend on the relative stiffness of the interface and the bulk, as well as on the geometry of the contact (which controls stress distribution in the near-surface region). It is possible that the wide range of values of activation volumes reported in literature for the same materials systems could arise from the experiments probing the interface and the bulk regions in varying degrees. This finding will be important to future studies of mechanochemical coupling as it suggests that the properties of the surrounding bulk must be taken into account. We have also found that, even for stiff materials, the near-surface region can be surprisingly compliant and therefore it may dominate the mechanochemical response.